\documentclass[preprint,preprintnumbers,amsmath,amssymb]{revtex4}

\usepackage{graphicx}
\usepackage{dcolumn}
\usepackage{bm}
\usepackage{rotating}

\begin{document}

\title[Conserved Hamiltonian in dissipative systems]
{Non-standard conserved Hamiltonian structures in
dissipative/damped systems : Nonlinear generalizations of damped
harmonic oscillator  }
\author{R Gladwin Pradeep, V K Chandrasekar, M Senthilvelan and M Lakshmanan}
\affiliation{Centre for Nonlinear Dynamics, School of Physics,
Bharathidasan University, Tiruchirappalli - 620 024, India }

\date{\today}

\newcommand{\ud}{\mathrm{d}}


\begin{abstract}
In this paper we point out the existence of a remarkable nonlocal
transformation between the damped harmonic oscillator and a
modified Emden type nonlinear oscillator equation with linear
forcing, $\ddot{x}+\alpha x\dot{x}+\beta x^3+\gamma x=0,$ which
preserves the form of the time independent integral, conservative
Hamiltonian and the equation of motion.  Generalizing this
transformation we prove the existence of non-standard conservative
Hamiltonian structure for a general class of damped nonlinear
oscillators including Li\'enard type systems.  Further, using the
above Hamiltonian structure for a specific example namely the
generalized modified Emden equation $\ddot{x}+\alpha
x^q\dot{x}+\beta x^{2q+1}=0$, where $\alpha$, $\beta$ and $q$ are
arbitrary parameters, the general solution is obtained through
appropriate canonical transformations.  We also present the
conservative Hamiltonian structure of the damped
Mathews-Lakshmanan oscillator equation. The associated Lagrangian
description for all the above systems is also briefly discussed.
\end{abstract}


\maketitle
\section{Introduction}
\label{sec1}
Dissipative systems are dynamical systems whose phase space volume
decreases/varies when the dynamical system evolves in time.
This should be contrasted with the conservative systems whose
 phase space volume remains  a constant.  A dynamical system,
which is represented by a second order ordinary differential equation (ODE) is said
to be a dissipative one if the flow function of the equivalent system of first order ODEs turns out
to be
 a negative constant$^{\footnotesize{1,2}}$.  In the case when
the flow function becomes zero the underlying system is by
definition a conservative one$^{\footnotesize{1,2}}$.  A typical
example is the damped harmonic oscillator (DHO) equation,
$\ddot{x}+\alpha \dot{x}+\lambda x=0$, where overdot denotes
differentiation with respect to time and $\alpha$ and $\lambda$
are arbitrary parameters.  Working out the flow function $\Lambda$
for this system one finds that $\Lambda=\frac{\partial
f_1}{\partial x} +\frac{\partial f_2}{\partial y}=-\alpha$, where
$f_1$ and $f_2$  are defined by
$\dot{x}=y=f_1(x,y);\,\dot{y}=-\alpha y-\lambda x=f_2(x,y)$, which
confirms that the system under consideration is a dissipative one.
On the other hand, for a conservative system, we have a set of
first order equations, $\dot{x}=\frac{\partial H}{\partial
p}=f_1(x,p);\, \dot{p}=-\frac{\partial H}{\partial x}=f_2(x,p)$,
where $H$ is the Hamiltonian function, and the flow function is
always equal to zero which can be straightforwardly confirmed by
substituting the Hamilton equations of motion.  This observation
reveals the fact that if one is able to find a time independent
integral for a dissipative system then this time independent
integral can be correlated to a Hamiltonian function, perhaps in a
non-standard form in terms of a new set of canonical variables, for
the second order equation under consideration.  Such a conserved
Hamiltonian description
 also leads one to further investigations, including quantization$^{\footnotesize{3}}$.

 Since the pioneering work of Bateman$^{\footnotesize{4}}$ the quest
for a Hamiltonian description for the damped harmonic oscillator
equation was pursued by several authors and only very recently conserved
Hamiltonian description
for this system$^{\footnotesize{5}}$ for all the three parametric regimes have  been constructed by three of us
.  A possible quantization has also been suggested.

In a parallel investigation we have also found that
another dissipative type system, namely the modified Emden type equation (MEE), also
admits time independent integral and a conservative Hamiltonian description$^{\footnotesize{6}}$.
A question which now naturally arises is whether the aforementioned systems are isolated examples
or there
exists a wider class of dissipative systems that admit conservative Hamiltonian description.
If so, how to
isolate and classify them?  A more detailed investigation in this direction reveals the fascinating fact that one can
map the damped harmonic oscillator equation onto the modified Emden type equation
through a nonlocal transformation.  Interestingly the time independent
integrals and the Hamiltonian description for the nonlinear system can also be derived
from the linear one by simply substituting the same nonlocal transformation at the appropriate
places.

 In this paper, after presenting the above novel results, we report a rather general
 transformation which can map the damped harmonic oscillator to a larger
 class of damped nonlinear systems that admit conservative Hamiltonian
description.  We present the general nonlinear oscillator equation
and explicit forms of the Hamiltonian.
The Hamiltonian forms obtained by this procedure  are
of non-standard (not equal to the standard potential plus kinetic energy) forms.
 Recently considerable interest has been shown in the classification
 nonlinear dissipative equations which admit non-standard conservative Hamiltonian structure
 $^{\footnotesize{7,8,9}}$.  We then
consider a specific nonlinear system, $\ddot{x}+\alpha x^q\dot{x}+\beta x^{2q+1}=0$, where
$q$, $\alpha$ and $\beta$ are arbitrary parameters, which is a natural generalization
of the DHO equation and the MEE
and discuss the dynamics/integrability of this equation in some detail.
Explicit
solution for this equation was
constructed for the specific
parametric choices$^{\footnotesize{10,11}}$ $\beta=\frac{\alpha^2}{(q+2)^2}$ and $\beta=\frac{\alpha^2}{4(q+1)}$.
For the former  choice of parameter we have
recently shown that this equation can be linearized to the free particle equation through a
generalized linearizing transformation$^{\footnotesize{12}}$.   Also the $q=1$ case (MEE) has
been completely integrated by using the underlying Hamiltonian structure$^{\footnotesize{6}}$ for all
values of the parameter $\alpha$ and $\beta$.  However, neither the Hamiltonian
structure nor explicit solutions
 for arbitrary choices of $q$, $\alpha$ and $\beta$ for this equation have been
reported and we present the results here.  Finally we also present the Hamiltonian
structure of the damped Mathews-Lakshmanan oscillator, whose undamped version
exhibits amplitude-dependent harmonic type oscillatory solutions$^{\footnotesize{13,14}}$.

The plan of the paper is as follows.  In section II we discuss the
time independent integrals and Hamiltonians of the two dissipative
systems, namely the DHO and MEE with linear forcing and deduce the
nonlocal transformation which interrelates each other.  In section
III we introduce a more general nonlocal transformation and
substitute it into the DHO equation and construct the nonlinear
generalizations of the DHO.  We also derive the associated time
independent integrals and Hamiltonian for this general nonlinear
oscillator equation.  In section IV, we consider as a specific
example, namely the generalized MEE, and discuss the Hamiltonian
structure
 and obtain its general solution by integrating the canonical equations of motion after suitable
canonical transformations.  In section V, we present as a second
example the time independent integral of motion and conservative
Hamiltonian structure of the damped Mathews-Lakshmanan oscillator.
In section VI, we briefly discuss the associated Lagrangian
description for all the above systems. Finally, we present our
conclusion in section VII.

\section{Hamiltonian structure of DHO and MEE}
In this section we briefly recall the Hamiltonian dynamics
associated with the two dissipative systems, namely the damped
harmonic oscillator and the modified Emden type equation, and then
show how they are interrelated.  Note that the former one is a
linear system and latter one is a nonlinear system.  One may
essentially consider the transformation to be introduced as a
linearizing transformation of the latter.
\subsection{Damped harmonic oscillator}
To start with let us consider the damped harmonic oscillator (using a different notation for convenience of
comparison)
\begin{eqnarray}
y''+\alpha y'+\lambda y=0,\qquad \left('=\frac{d}{d\tau}\right)\label{damp}
\end{eqnarray}
where $\alpha$ and $\lambda$ are arbitrary parameters.
Recently, we have identified
the following time independent integral
of motion$^{\footnotesize{5}}$ for the system (\ref{damp}) :
\begin{eqnarray}
\hspace{-2cm}I=\left \{
\begin{array}{ll}
\displaystyle\frac{(r-1)}{(r-2)}\left(y'+\frac{\alpha}{r}y\right)(y'+\frac{(r-1)}{r}\alpha y)^{(1-r)},&\alpha^2>4\lambda\\\\
\displaystyle\frac{y'}{(y'+\frac{1}{2}\alpha y)}-\log[y'+\frac{1}{2}\alpha y],&\alpha^2=4\lambda\\\\
\displaystyle\frac{1}{2}\log[y'^2+\alpha y y'+\lambda y^2]+\frac{\alpha}{2\omega}
\displaystyle\tan^{-1}\left[\frac{\alpha y'+2\lambda y}{2\omega y'}\right],&\alpha^2<4\lambda,
\end{array}
\right.\label{damp_integral}
\end{eqnarray}
where $r=\frac{\alpha}{2\lambda}(\alpha\pm\sqrt{\alpha^2-4\lambda})$ and
 $\omega=\frac{1}{2}\sqrt{4\lambda-\alpha^2}$, for the overdamped, critically damped
and the underdamped oscillations, respectively.
From these time independent integrals we have also derived the following time independent Hamiltonian
 $^{\footnotesize{5}}$,
\begin{eqnarray}
H=\left\{
\begin{array}{ll}
\displaystyle\frac{(r-1)}{(r-2)}(p)^{\frac{(r-2)}{(r-1)}}-\frac{(r-1)}{r}\alpha p y,\qquad&p>0,\,\alpha^2>4\lambda\\\\
\displaystyle\log(p)-\frac{1}{2}\alpha p
y,&p>0,\,\alpha^2=4\lambda\\\\\label{damp_ham}
\displaystyle\frac{1}{2}\log[y^2\sec^2(\omega p y
)]-\frac{\alpha}{2}p y ,&\qquad\quad\alpha^2<4\lambda
\end{array}
\right.
\end{eqnarray}
where the canonical conjugate momentum $p$ is defined by
\begin{eqnarray}
p=\left \{
\begin{array}{ll}
\displaystyle\left(y'+\frac{(r-1)}{r}\alpha y\right)^{(1-r)},\qquad\qquad\quad&\alpha^2\ge4\lambda\\\\
\displaystyle\frac{1}{\omega y}\tan^{-1}\left[\frac{2y'+\alpha y}{2\omega y}\right].&\alpha^2<4\lambda
\end{array}
\right.\label{damp_mom}
\end{eqnarray}
From the Hamiltonian (\ref{damp_ham})-(\ref{damp_mom}) one can straightforwardly write down the canonical
equations of motion for all the three regimes and obtain the known exact solutions straightforwardly.
Note that no multivaluedness arises in the system due to the constraints on the momentum.\\
{\bf Case 1,2  $\alpha^2\ge4\lambda$}\\
Substituting the first of the expressions in (\ref{damp_ham}) into the canonical equations
$y'=\frac{\partial H}{\partial p}$
$p'=-\frac{\partial H}{\partial y}$, one gets the following equivalent system
of first order ordinary differential equations for the damped harmonic oscillator in the overdamped and critically
 damped ($\alpha^2>4\lambda$ and $\alpha^2=4\lambda$) parametric
regimes,
\begin{eqnarray}
y'=(p)^{\frac{1}{(1-r)}}-\frac{(r-1)}{r}\alpha y,\quad
p'=\frac{(r-1)}{r}\alpha p.\label{damp_can1}
\end{eqnarray}
One can easily check that the second order equivalence of this equation
coincides exactly with (\ref{damp}) and that the standard form of the solutions
for the overdamped and the critically damped cases follow naturally by integrating
the above system of first order ODEs.
 We mention here that the divergence of
the flow function of the underlying canonical equations of motion is zero, $\Lambda=\frac{\partial f_1}{\partial y}
+\frac{\partial f_2}{\partial p}=0$, which
in turn confirms that the damped harmonic oscillator has a non-standard conservative
Hamiltonian description.\\
{\bf Case 3 $\alpha^2<4\lambda$}\\
Finally, for the underdamped case, one can write down the Hamilton equations of motion
in the form
\begin{eqnarray}
\displaystyle y'=\frac{1}{2}y(2\omega\tan(\omega p y)-\alpha),\,\,\,
p'=\frac{1}{2}p(\alpha-2\omega\tan(\omega py))-\frac{1}{y}.\label{damp_can2}
\end{eqnarray}
Again rewriting this equation as a single second order equation in the variable $y$
one ends up in the form (\ref{damp}).  The divergence of the flow function
in this case is also zero.

The existence of time independent Hamiltonians  (\ref{damp_ham})
in a non-standard form and the basic definition of the divergence
of flow function lead us to the conclusion that the damped
harmonic oscillator has a non-standard conservative Hamiltonian
description as well, in spite of its known dissipative nature.
\subsection{Modified Emden type equation}
In a parallel investigation we have also found that the MEE, $\ddot{x}+\alpha x\dot{x}+\beta x^3=0$,  admits a
time independent
conservative Hamiltonian description$^{\footnotesize{6}}$.  The MEE,
which is also known as the Painlev\'e - Ince equation, is an extensively studied and
physically significant equation in the contemporary nonlinear dynamics
literature$^{\footnotesize{15-19}}$.
 On the other
hand the MEE with linear external forcing,
\begin{eqnarray}
\ddot{x}+\alpha x\dot{x}+\beta x^3+\gamma x=0,\label{linearforce}
\end{eqnarray}
possesses certain unusual nonlinear dynamical properties
 for the parametric choice $\beta=\frac{\alpha^2}{9}$.
We have shown that this system admits nonisolated periodic orbits of conservative Hamiltonian type$^{\footnotesize{9}}$. These
periodic orbits
exhibit the unexpected property that the frequency of oscillations is completely independent of amplitude and
continues to remain as that of the linear harmonic oscillator$^{\footnotesize{9}}$.

The more general system (\ref{linearforce}) with $\alpha$, $\beta$ and $\gamma$ being arbitrary parameters also
admits the following time independent
integrals of motion$^{\footnotesize{21}}$,
\begin{eqnarray}
\hspace{-2cm}I=\left \{
\begin{array}{ll}
\displaystyle\frac{(r-1)}{(r-2)}\left(\dot{x}+\frac{\alpha}{2r}x^2+\frac{r\gamma}{\alpha(r-1)})
\right)\bigg(\dot{x}+\frac{\alpha (r-1)}{2r}
x^2+\frac{r\gamma}{\alpha}\bigg)^{(1-r)},&\alpha^2>8\beta\\\\
\displaystyle\frac{\dot{x}}{(\dot{x}+\frac{1}{4}\alpha(x^2+\frac{4\gamma}{\alpha^2}))}
-\log[\dot{x}+\frac{1}{4}\alpha (x^2+\frac{4\gamma}{\alpha^2})],&\alpha^2=8\beta\\\\
\displaystyle\frac{1}{2}\log[\dot{x}^2+\frac{\alpha}{2} \dot{x}(x^2+\frac{\gamma}{\beta})
+\frac{\beta}{2} (x^2+\frac{\gamma}{\beta})^2]\\
\qquad+\frac{\alpha}{2\omega}
\displaystyle\tan^{-1}\left[\frac{\alpha \dot{x}+2\beta (x^2+\frac{\gamma}{\beta})}
{2\omega\dot{x}}\right],&\alpha^2<8\beta,
\end{array}
\right.\label{linearforceint}
\end{eqnarray}
where $r=\frac{\alpha}{4\beta}(\alpha\pm\sqrt{\alpha^2-8\beta})$ and
 $\omega=\frac{1}{2}\sqrt{8\beta-\alpha^2}$.  From the time independent integrals we have identified
the following Hamiltonian for the system (\ref{linearforce}),
\begin{eqnarray}
\hspace{-0.5cm}H=\left\{
\begin{array}{ll}
\displaystyle\frac{(r-1)}{(r-2)}(p)^{\frac{(r-2)}{(r-1)}}-\frac{(r-1)}{2r}\alpha px^2
-\frac{pr\gamma}{\alpha},\qquad&\alpha^2>8\beta\\\\
\displaystyle\log(p)-\frac{1}{2}\alpha p(\frac{x^2}{2}+\frac{4\gamma}{\alpha^2}),
&\alpha^2=8\beta\\\\
\displaystyle\frac{1}{2}\log[(x^2+\frac{\gamma}{\beta})^2\sec^2(\frac{\omega  p}{2}
(x^2+\frac{\gamma}{\beta}))]-\frac{\alpha}{4} p (x^2+\frac{\gamma}{\beta}),
&\alpha^2<8\beta
\end{array}
\right.\label{linearforceh}
\end{eqnarray}
where the canonical conjugate momentum $p$ is defined by
\begin{eqnarray}
\hspace{-0.5cm}p=\left \{
\begin{array}{ll}
\displaystyle\left(\dot{x}+\frac{(r-1)}{2r}\alpha x^2+\frac{r\gamma}{\alpha}
\right)^{(1-r)},\qquad\qquad\quad&\alpha^2\ge8\beta\\\\
\displaystyle\frac{2}{\omega (x^2+\frac{\gamma}{\beta})}
\tan^{-1}\left[\frac{4\dot{x}+\alpha (x^2+\frac{\gamma}{\beta})}
{2\omega (x^2+\frac{\gamma}{\beta})}\right].&\alpha^2<8\beta
\end{array}\label{linearforcep}
\right.
\end{eqnarray}

The Hamilton equations of motion follows straightaway from (\ref{linearforceh})
as
\begin{eqnarray}
&&\hspace{-2.5cm}\mbox{\emph{Cases} 1 \& 2}\hspace{0.5cm}\dot{x}=p^{\frac{1}{1-r}}
-\frac{\alpha(r-1)x^2}{2r}-\frac{r\gamma}{\alpha},\quad
\dot{p}=\frac{\alpha(r-1)px}{r}\label{linearforceeq1}\\
&&\hspace{-2.5cm}\mbox{\emph{Case} 3}\hspace{1.4cm}\dot{x}=\frac{1}{4\beta}(\beta x^2+\gamma)\left(2\omega\tan\left[\frac{1}{2}p\omega(x^2+\frac{\gamma}{\beta})\right]
-\alpha\right),\nonumber\\
&&\hspace{0.2cm}\dot{p}=\frac{x(p\alpha(\beta x^2+\gamma)-4\beta
-2p\omega (\beta x^2+\gamma)
\tan[\frac{1}{2}p\omega(x^2+\frac{\gamma}{\beta})])}{2(\beta x^2+\gamma)}.\label{linearforceeq3}
\end{eqnarray}
Note that in all the above three cases, the divergence of the flow function $\Lambda$ is zero, corresponding to
a a conservative Hamiltonian description.  However, unlike the damped harmonic oscillator one cannot integrate the canonical
equations (\ref{linearforceeq1}) and (\ref{linearforceeq3}) and obtain the solutions straightforwardly.
To overcome this difficulty
one should introduce suitable canonical transformations and change the Hamilton equations (\ref{linearforceeq1}) and (\ref{linearforceeq3})
into simpler
forms so that they can be integrated in a straightforward manner as demonstrated in the case of MEE
(with $\gamma=0$) recently by us$^{\footnotesize{6}}$.

\subsection{Transformation connecting DHO and MEE}
By comparing the structure of the integral of motion, Hamiltonian function and canonical equations of the damped
harmonic oscillator
with that of the  equation (\ref{linearforce}), namely equation (\ref{damp_integral}) with (\ref{linearforceint}),
(\ref{damp_ham})-(\ref{damp_mom})
with (\ref{linearforceh})-(\ref{linearforcep}) and (\ref{damp_can1})-(\ref{damp_can2}) with
(\ref{linearforceeq1})-(\ref{linearforceeq3}), one
can identify that the two systems are transformed into each other through the nonlocal transformation
\begin{eqnarray}
y=\frac{x^2}{2}+\frac{\gamma}{\lambda},\quad d\tau=xdt,\label{linearforcet}
\end{eqnarray}
with the identification $\lambda=2\beta$.
One can also check directly that the above transformation also maps the damped
harmonic oscillator equation (\ref{damp}) onto the MEE with forcing, equation (\ref{linearforce}), and vice versa.
Consequently one can treat the transformation as a linearizing transformation (albeit nonlocal) of the
nonlinear equation (\ref{linearforce}).

Note that the transformation (\ref{linearforcet}) is not the only possible linearizing transformation at least
for specific parametric choices.  We mention here that one can also transform the nonlinear system (\ref{linearforce})
with $\beta=\frac{\alpha^2}{9}$ into a linear harmonic oscillator
equation $(\ddot{U}+\gamma U=0)$ by introducing another nonlocal
transformation of the form $^{\footnotesize{9,22}}$
\begin{eqnarray}
U=xe^{\int_0^t \alpha x(\hat{t})d\hat{t}},\,\,\,\tau=t.\label{utrans}
\end{eqnarray}
The nonlocal transformation (\ref{utrans}) is different from (\ref{linearforcet})
in the following respect.  In (\ref{linearforcet}) the nonlocality is introduced
in the independent variable whereas in (\ref{utrans}) the nonlocality is introduced
in the dependent variable.  We may also add that the nonlocal transformation (\ref{linearforcet}) has
some similarity with the well known Kustaanheimo-Stiefel transformation used in atomic physics$^{\footnotesize{23}}$.
Even though both the nonlocal transformations (\ref{linearforcet}) and (\ref{utrans}) map
the nonlinear equation into a linear one and vice versa, the nonlocal transformation of the type
(\ref{linearforcet}) is much useful in identifying Hamiltonian structures associated
with the nonlinear system whereas the nonlocal transformation of the type (\ref{utrans}) is
more useful in constructing general solution for the transformed nonlinear system.
For more details about the nonlocal transformation of the type (\ref{utrans}) we
may refer to Ref. 22.  In the following we confine our attention to the nonlocal
transformation of the form (\ref{linearforcet}) and its generalization.
\section{A General class of nonlinear damped oscillator : Hamiltonian description}
We find that the transformation (\ref{linearforcet}) is a specific case of a rather general nonlocal
transformation of the form
\begin{eqnarray}
y=\int f(x)dx,\quad  d\tau=\frac{f(x)}{g(x)}dt.\label{nlt}
\end{eqnarray}
For example, restricting $f(x)=x$ and $g(x)=1$ in (\ref{nlt}) one gets exactly (\ref{linearforcet}).

The nonlocal transformation (\ref{nlt})
modifies the damped harmonic oscillator equation (\ref{damp}) to the general class of nonlinear
oscillators of the form,
\begin{eqnarray}
\ddot{x}+\frac{g'(x)}{g(x)}\dot{x}^2+\alpha \frac{f(x)}{g(x)}\dot{x} +\lambda \frac{f(x)}{g(x)^2}\int f(x)dx=0,\label{lienard}
\end{eqnarray}
where $f(x),\,g(x)$ are arbitrary function of $x$.

Applying the above nonlocal transformation (\ref{nlt}) to the damped harmonic oscillator equation (\ref{damp})
we obtain the following time-independent integral of motion for the
nonlinear system (\ref{lienard}), that is,    \\
\noindent Case 1. $\alpha^2>4\lambda$
\begin{eqnarray}
I=\frac{(r-1)}{(r-2)}\left(g(x)\dot{x}+\frac{\alpha}{r}\int f(x)dx\right)
\left(g(x)\dot{x}+\frac{(r-1)}{r}\alpha\int f(x)dx\right)^{(1-r)}\label{eqint1}
\end{eqnarray}
Case 2. $\alpha^2=4\lambda$
\begin{eqnarray}
I=\frac{g(x)\dot{x}}{g(x)\dot{x}+\frac{1}{2}\alpha\int f(x)dx}-\log[g(x)\dot{x}+\frac{1}{2}\alpha\int f(x)dx]
\end{eqnarray}
 Case 3. $\alpha^2<4\lambda$
\begin{eqnarray}
&&I=\frac{1}{2}\log[g(x)^2\dot{x}^2+\alpha g(x)\dot{x}\int f(x)dx+\lambda (\int f(x) dx)^2]\nonumber\\
&&\qquad+\frac{\alpha}
{2 \omega}\tan^{-1}\left[\frac{\alpha g(x)\dot{x}+2\lambda \int f(x)dx}{2\omega g(x) \dot{x}}\right],
\label{eqint3}
\end{eqnarray}
\vskip 5pt
\noindent where $\omega=\frac{1}{2}\sqrt{4\lambda-\alpha^2}$ and
$r=\frac{\alpha}{2\lambda}(\alpha\pm\sqrt{\alpha^2-4\lambda})$.

The Hamiltonian for the equation (\ref{lienard}) can also be constructed by
simply substituting the transformation (\ref{nlt}) into (\ref{damp_ham})
so that the latter reads
\vskip 4pt
\begin{eqnarray}
\hspace{-1.8cm}H=\left \{
\begin{array}{ll}
\displaystyle\frac{(r-1)}{(r-2)}\left(\frac{p}{g(x)}\right)^{\frac{r-2}{r-1}}
-\frac{\alpha(r-1)}{r}\frac{p}{g(x)} \int f(x)dx,&\alpha^2>4\lambda\\\\
\displaystyle\frac{\alpha}{2g(x)}p\int f(x)dx+\log\left[\frac{g(x)}{p}\right],& \alpha^2=4\lambda\\\\
\displaystyle\frac{1}{2}\log\left[(\int f(x)dx)^2
\sec^2\left[\frac{\omega p}{g(x)}\int f(x)dx\right]\right]
-\frac{\alpha p}{2g(x)}\int f(x)dx,& \alpha^2<4\lambda,
\end{array}
\right.\label{lienard_ham}
\end{eqnarray}
\noindent where $p$ is the canonical conjugate momentum defined by
\begin{eqnarray}
p=\left\{\displaystyle
\begin{array}{ll}
\displaystyle\frac{g(x)}{\left[g(x)\dot{x}+\frac{(r-1)}{r}
\alpha\int f(x)dx\right]^{r-1}},&\alpha^2\ge4\lambda\\\\
\displaystyle\frac{g(x)}{\omega\int f(x)dx}
\tan^{-1}\left[\frac{2g(x)\dot{x}+\alpha\int f(x)dx}{2\omega\int
f(x)dx}\right], &\alpha^2<4\lambda.\label{lienard_mom}
\end{array}
\right.
\end{eqnarray}
\vskip 4pt
\noindent One can easily check that the canonical equations of motion take the form
\begin{eqnarray}
&&\hspace{-2.0cm}\alpha^2\ge4\lambda\,\,:\hspace{0.1cm}\dot{x}=\frac{1}{g(x)}
\left[\left(\frac{p}{g(x)}\right)^{\frac{1}{1-r}}-\frac{(r-1)}{r}
\alpha\int f(x)dx\right],\nonumber\\
&&\hspace{-0.2cm}
\dot{p}=\frac{(r-1)}{r}\alpha p \frac{f(x)}{g(x)}+
\frac{pg'(x)}{g(x)^2}\left[\left(\frac{p}{g(x)}\right)^{\frac{1}{1-r}}
-\frac{\alpha(r-1)}{r}\int f(x)dx\right].\label{can1}\\
&&\hspace{-2.0cm}\alpha^2<4\lambda\,\,:\hspace{0.1cm}
\dot{x}=\frac{\int f(x)dx}{2g(x)}\left[2\omega\tan
\left[\frac{p\omega}{g(x)}\int f(x) dx\right]-\alpha\right],\nonumber\\
&&\hspace{-0.2cm}\dot{p}=\frac{1}{2g(x)^2\int f(x)dx}
\bigg[f(x)g(x)\bigg(2g(x)-\bigg(\alpha-2\omega\tan\left[\frac{p\omega\int f(x)dx}{g(x)}
\right]\bigg)\nonumber\\
&&\hspace{0.2cm}\times p\int f(x)dx\bigg)+p\left(\int f(x)dx\right)^2\left
(\alpha-2\omega\tan\left[\frac{p\omega\int f(x)dx}{g(x)}\right]
\right)g'(x)\bigg].\label{can3}
\end{eqnarray}
 One can straightforwardly check that the second order equivalence
of the equations (\ref{can1}) and (\ref{can3}) coincides exactly with (\ref{lienard}) in the appropriate parametric regimes.
Further, we find that the flow function for the  canonical equations of motion
is zero in all the three parametric regions, equations (\ref{can1})-(\ref{can3}), which in turn confirms that
(\ref{lienard}) or
(\ref{can1})-(\ref{can3}) has a Hamiltonian
description.  Further, the integrability of (\ref{lienard})  is automatically ensured
by the existence of the time independent Hamiltonian (\ref{lienard_ham}) in the
 Liouville sense.

We also note here that equation (\ref{lienard}) can be transformed into an Abel equation of the second
kind$^{\footnotesize{24}}$,
\begin{eqnarray}
ww'+\frac{g'(x)}{g(x)}w^2+\alpha \frac{f(x)}{g(x)}w+\lambda\frac{f(x)}{g(x)^2}\int f(x)dx=0,
\label{abel}
\end{eqnarray}
through the transformation $w(x)=\dot{x}$.  Then the time
independent
first integrals of (\ref{abel}) with $g(x)=$  constant
can be constructed by the procedure described in Ref. 25.
In the case where $\int f(x)dx$ is an invertible function of $x$, then the
procedure given in Ref. 25 to construct the solution will lead
 to quadratures in terms of certain complicated integrals which however cannot be evaluated in general.
For further details one may refer to Ref. 26.
Therefore one should adopt a different procedure to construct the general solution of
(\ref{lienard}).
The effective way to proceed further is to
 transform the Hamiltonian (\ref{lienard_ham}) to a simpler form
 through suitable canonical transformations as demonstrated recently by us for the case of the MEE$^{\footnotesize{6}}$.
 However, to adopt this
procedure one should specify the explicit forms of $f(x)$ and $g(x)$.  We demonstrate the procedure with a
specific example in the following section.

\section{Generalized MEE : Hamiltonian
structure and general solution}
To illustrate the ideas given in section 3,  we focus our attention on the case
$f(x)=x^q$, $g(x)=1$ so that equation (\ref{lienard}) now becomes
\begin{eqnarray}
\ddot{x}+\alpha x^q \dot{x}+\beta x^{2q+1}=0,\label{eqq}
\end{eqnarray}
where $\beta=\frac{\lambda}{(q+1)}$.
The reason for choosing this form of $f(x)$ for illustration is that it provides a natural
generalization of the damped harmonic oscillator and the modified Emden
type equation.


\label{sec2}

Substituting
 $f(x)=x^q$
in equations (\ref{eqint1}) - (\ref{eqint3}) we obtain the following time
independent integral of motion for the equation (\ref{eqq}), that is
\vskip 4pt
\noindent Case 1. $\alpha^2>4\beta(q+1)$
\begin{eqnarray}
I=\frac{(r-1)}{(r-2)}\left(\dot{x}+\frac{\hat{\alpha}}{r}x^{q+1}\right)
\left(\dot{x}+\frac{(r-1)}{r}\hat{\alpha} x^{q+1}\right)^{(1-r)}\label{gen_emden_int1}
\end{eqnarray}
Case 2. $\alpha^2=4\beta(q+1)$
\begin{eqnarray}
I=\frac{\dot{x}}{\dot{x}+\frac{\hat{\alpha}}{2}x^{q+1}}-\log[\dot{x}+\frac{\hat{\alpha}}{2}x^{q+1}]
\end{eqnarray}
 Case 3. $\alpha^2<4\beta(q+1)$
\begin{eqnarray}
&&\hspace{-1cm}I=\frac{1}{2}\log[\dot{x}^2+\hat{\alpha} \dot{x}x^{q+1}+\frac{\beta}{(q+1)} x^{2(q+1)}]
+\frac{\alpha}{2 \omega}\tan^{-1}\left[\frac{\alpha \dot{x}+2\beta x^{q+1}}{2\omega \dot{x}}\right],
\label{gen_emden_int3}
\end{eqnarray}
where $r=\frac{\alpha}{2\beta(q+1)}(\alpha\pm\sqrt{\alpha^2-4\beta(q+1)})$,
$\omega=\frac{1}{2}\sqrt{4\beta(q+1)-\alpha^2}$
and $\hat{\alpha}=\frac{\alpha}{q+1}$.
The Hamiltonian for the equation (\ref{eqq}) can be deduced from equation (\ref{lienard_ham}) which
turns out to be
\vskip 4pt
\begin{eqnarray}
H=\left\{\label{emden_ham}
\begin{array}{ll}
\displaystyle\frac{(r-1)}{(r-2)}p^{\frac{(r-2)}{(r-1)}}
-\frac{(r-1)}{r} \hat{\alpha} px^{q+1}, & \alpha^2>4\beta (q+1)\\\\
\displaystyle\log(p)-\frac{\hat{\alpha}}{2}px^{q+1},&\alpha^2=4\beta (q+1)\\\\
\displaystyle\frac{\hat{\alpha}}{2}px^{q+1}-\frac{1}{2}\log\left[x^{2(q+1)}
\sec^2[\frac{\omega}{(q+1)}x^{q+1}p]\right],&\alpha^2<4\beta (q+1),
\end{array}
\right.
\end{eqnarray}
\noindent where the canonically conjugate momentum is defined by
\begin{eqnarray}
p=\left\{
\begin{array}{ll}
\displaystyle\frac{1}{(r-1)}\left(\dot{x}+\frac{(r-1)}{r}
\hat{\alpha}x^{q+1}\right)^{(1-r)},\\\\
\displaystyle\frac{(q+1)}{\omega x^{q+1}}\tan^{-1}\left[\frac{\alpha x^{q+1}
+2(q+1)\dot{x}}{2\omega x^{q+1}}\right].
\end{array}
\right.
\end{eqnarray}
\vskip 4pt
Here we note that the integrals (\ref{gen_emden_int1})-(\ref{gen_emden_int3}) can also be derived systematically through
various methods available in the recent literature.  To name a few, we cite Prelle-Singer
procedure$^{\footnotesize{27}}$,  method of reducing Li\'enard equation to the Abel equation form$^{\footnotesize{25}}$, factorization
method$^{\footnotesize{28,29}}$ and so on.  However, all these methods provide only time independent integrals
but neither the integrals nor the Hamilton canonical equations can be integrated
straightforwardly.

\subsection{Method of obtaining general solution}
\label{sec4}
In this section we transform the Hamiltonian (\ref{emden_ham}) into a simpler form,  by introducing a suitable canonical
transformation and then obtain the solutions.\\\\
{\bf Case 1: $\alpha^2>4\beta (q+1)$}
\vskip 4pt
\noindent The Hamiltonian for this parametric regime is given by
\begin{eqnarray}
H=\frac{(r-1)}{(r-2)}p^{\frac{(r-2)}{(r-1)}}
-\frac{(r-1)}{r}\hat{\alpha} p x^{q+1},\,\,r=\frac{(\alpha^2\pm\alpha\sqrt{\alpha^2-4\beta(q+1)})}{2\beta(q+1)}\hspace{0.5cm}
(\ref{emden_ham})\nonumber
\end{eqnarray}
By introducing a canonical transformation
\begin{eqnarray}
x=q\left(\frac{U}{P}\right)^{\frac{1}{q}},\quad
p=\frac{1}{2}\left(U^{q-1}P^{q+1}\right)^{\frac{1}{q}}\label{canon}
\end{eqnarray}
the Hamiltonian (\ref{emden_ham}) can be transformed into the
 form
\begin{eqnarray}
H=\sigma_1P^{n_1}U^{m_1}+\eta_1U^2=E,\label{can_ham1}
\end{eqnarray}
where we have defined $\sigma_1=\frac{(r-1)}{(r-2)}\left(\frac{1}{2}\right)^{\frac{(r-2)}{(r-1)}}$,
$n_1=\frac{(q+1)(r-2)}{q(r-1)}$, $m_1=\frac{(q-1)(r-2)}{q(r-1)}$ and
$\eta_1=\frac{(1-r)}{2r}\hat{\alpha}q^{(q+1)}$.\\

The canonical equations of motion for the transformed Hamiltonian
(\ref{can_ham1}) now reads
\begin{subequations}
\begin{eqnarray}
\addtocounter{equation}{-1}
 \label{can_eq1}
\addtocounter{equation}{1}
&&\dot{U}=\sigma_1n_1U^{m_1}P^{n_1-1},\label{can_eq1a}\\
&& \dot{P}=-(\sigma_1m_1P^{n_1}U^{m_1-1}+2\eta_1U).\label{can_eq1b}
\end{eqnarray}
\end{subequations}
 One may observe that for the choice $n_1=1$
 equation (\ref{can_eq1}) becomes uncoupled and one can integrate the resultant equations straightforwardly.
In the following, first we
consider the general case and then discuss the particular case $n_1=1$ which corresponds to the
parametric choice $\beta=\frac{\alpha^2}{(q+2)^2}$.

Substituting  (\ref{can_eq1a}) into (\ref{can_ham1}) and rewriting the latter, we get
\begin{eqnarray}
E&&=\frac{\sigma_1U^{m_1}}{(\sigma_1n_1)^{\frac{n_1}{n_1-1}}}
\left[\dot{U}^{\frac{n_1}{n_1-1}}U^{\frac{m_1n_1}{1-n_1}}\right]+\eta_1U^2.
\label{eqE}
\end{eqnarray}
Rewriting equation (\ref{eqE}) we obtain
\begin{eqnarray}
\dot{U}&&=\left[\frac{1}{\hat{\sigma}_1}[E-\eta_1U^2]\right]^{\frac{n_1-1}{n_1}}
U^{\frac{m_1}{n_1}}, \qquad n_1\ne 1
\end{eqnarray}
where $\hat{\sigma}_1=\frac{\sigma_1}{(\sigma_1n_1)^{\frac{n_1}{n_1-1}}}$.
 Integrating the above equation one obtains,
\begin{eqnarray}
t-t_0=&&\int\frac{\hat{\sigma}_1^nU^{m}dU}{[E-\eta_1U^2]^{n}},
\end{eqnarray}
where $m=\frac{-m_1}{n_1}$ and $n=\frac{n_1-1}{n_1}$.
The above integral can be split into two cases depending upon the values of $m$
and $n$, that is
\begin{eqnarray}
\hspace{-1.0cm}
t-t_0=\left \{
\begin{array}{ll}
\displaystyle
 \frac{U^{m-1}}{\eta_1(2n-m-1)(E-\eta_1U^2)^{n-1}}\\
\displaystyle\hspace{4cm}+\frac{(m-1)E}{\eta_1(2n-m-1)}\int\frac{U^{m-2}}{(E-\eta_1U^2)^n}dU
&  m,n>0
\\\\\label{quadratures}
\displaystyle
 \frac{1}{E(1-m)U^{m-1}(E-\eta_1U^2)^{n-1}}\\
\displaystyle\hspace{4cm}+\frac{\eta_1(m+2n-3)}{E(1-m)}\int\frac{1}{U^{m-2}(E-\eta_1U^2)^n}dU
&  m<0.
\end{array}
\right.
\end{eqnarray}
The integrals on the right hand sides of (\ref{quadratures}) can be integrated again and again
until all the integrals are exhausted, thereby giving the general solution in an implicit form.

Now we analyze the case $n_1=1$ in (\ref{can_eq1}).  For this choice, equation (\ref{can_eq1a}) can be integrated
to yield $U$
in the form
\begin{eqnarray}
U=[(1-m_1)(\sigma_1t+c_1)]^{\frac{1}{1-m_1}},\label{U}
\end{eqnarray}
where $c_1$ is an integration constant.
From equation (\ref{can_ham1}) one can express $P$ in terms of $U$ and $E$ and
substituting the latter into the first expression in (\ref{canon}) one gets
\begin{eqnarray}
x=q\left(\frac{\sigma_1U^{m_1+1}}{E-\eta_1U^2}\right)^{\frac{1}{q}},\label{xinu}
\end{eqnarray}
where $\sigma_1=\frac{q+1}{q}\left(\frac{1}{2}\right)^{\frac{q+1}{q}}$,
 $\eta_1=-\alpha\frac{q^{q+1}}{2(q+2)}$ and $m_1=\frac{q-1}{q+1}$.  Substituting
 (\ref{U}) into (\ref{xinu}) and simplifying
the resultant expression we arrive at
\begin{eqnarray}
x(t)=\left(\frac{2^{\frac{q^2-1}{q}}q^{q-1}(q+1)^2(q+2)(\sigma_1t+c_1)^q}
{2E(q+2)+\alpha2^{q+1}q^{q+1}(\sigma_1t+c_1)^{q+1}}\right)^{\frac{1}{q}}.\label{solution}
\end{eqnarray}
with $E$ and $c_1$ are two arbitrary constants.
We mention here that the linearizable case $\alpha^2=9\beta$ belongs to this case
and the solution
exactly agrees with the known one in the literature$^{\footnotesize{10-12}}$.  \\

\noindent{\bf Case 2 : $\alpha^2=4\beta(q+1)$}
\vskip 4pt
Using the same canonical transformation, equation (\ref{canon}), one can transform
the Hamiltonian,
$\displaystyle H=\log(p)-\frac{\hat{\alpha}}{2}px^{q+1}$,
into the form
\begin{eqnarray}
H=\frac{(q-1)}{q}\log(U)+\frac{q+1}{q}\log[P]+\eta_2 U^2=E\label{ham2}
\end{eqnarray}
where $\eta_2=-\frac{\hat{\alpha}}{4}q^{(q+1)}$.
The corresponding canonical
equations become
\begin{subequations}
\begin{eqnarray}
\addtocounter{equation}{-1}
 \label{can_eq2}
\addtocounter{equation}{1}
&&\dot{U}=\frac{(q+1)}{qP},\label{can_eq2a}\\
&& \dot{P}=\left(\frac{1-q}{q}\right)\frac{1}{U}+2\eta_2U.\label{can_eq2b}
\end{eqnarray}
\end{subequations}
From (\ref{can_eq2a}) one can express $P$ in terms of  $\dot{U}$ and substituting
this into (\ref{ham2}) one can bring the latter to the form
\begin{eqnarray}
\hat{E}=\frac{(q-1)}{q}\log(U)-\frac{(q+1)}{q}\log(\dot{U})+\eta_2U^2
\end{eqnarray}
which in turn gives us
\begin{eqnarray}
\dot{U}=\mbox{exp}[\frac{q\eta_2U^2}{(q+1)}]U^{\frac{q-1}{q+1}}E_1,\label{udot}
\end{eqnarray}
where $E_1=\mbox{exp}[\frac{-q\hat{E}}{(q+1)}]$.  Integrating the above
equation, one obtains the solution in the form
\begin{eqnarray}
t-t_0&=&\frac{1}{E_1}\int\frac{dU}{\mbox{exp}[\frac{q\eta_2U^2}{(q+1)}]
U^{\frac{(q-1)}{q+1}}}\nonumber\\
&=&-\frac{1}{2E_1}U^{\frac{2}{q+1}}
\left(\frac{q+1}{q\eta_2U^2}\right)^{\frac{1}{q+1}}\Gamma\left[\frac{1}{q+1},
\frac{q\eta_2U^2}{q+1}\right],\label{quad1}
\end{eqnarray}
where $\Gamma$ is the gamma function$^{\footnotesize{30}}$.  For the choice $q=1$ the integral
(\ref{quad1}) can be evaluated in terms of error function $^{\footnotesize{6,30}}$.\\\\
{\bf Case 3 : $\alpha^2<4\beta(q+1)$}
\vskip 4pt
Now we focus our attention on the underdamped case.
Using the canonical transformation (\ref{canon}), we rewrite
the underlying Hamiltonian
in the form
\begin{eqnarray}
H=\frac{\hat{\alpha}}{4}q^{(q+1)}U^2-
\log\left[\left(\frac{U}{P}\right)^{\frac{(q+1)}{q}}
\sec[\frac{\omega}{2(q+1)}U^2]\right].\label{ham3}
\end{eqnarray}
The associated canonical equations read
\begin{subequations}
\begin{eqnarray}
\addtocounter{equation}{-1}
 \label{udot1}
\addtocounter{equation}{1}
&&\dot{U}=\frac{q+1}{qP},\label{udot1a}\\
&&\dot{P}=\frac{2q\omega U^2\tan[\frac{\omega U^2}{2(q+1)}]
-((q+1)q^{q+2}U^2\alpha-2(q+1)^2)}{2q(q+1)U}.\label{udot1b}
\end{eqnarray}
\end{subequations}
From (\ref{udot1a}) one can write $P=\frac{q+1}{q\dot{U}}$ and substituting this
in the Hamiltonian (\ref{ham3}) and simplifying we get
\begin{eqnarray}
\tilde{H}=\log[U\dot{U}^{\frac{q+1}{q}}\sec[\frac{\omega}{2(q+1)}U^2]]+\eta_2U^2=E.
\label{tildeh}
\end{eqnarray}
  Rewriting equation (\ref{tildeh}) we have
\begin{eqnarray}
\dot{U}=\frac{1}{U}\frac{\mbox{exp}[\frac{q}{q+1}(E-\eta_2U^2)]}
{\sec[\frac{\omega}{2(q+1)}U^2]^{\frac{q}{q+1}}}.\label{udot2}
\end{eqnarray}
Integrating equation (\ref{udot2}), we get
\begin{eqnarray}
\hspace{-1cm}t-t_0&=&\int\frac{U\sec[\frac{\omega}{2(q+1)}U^2]^{\frac{q}{q+1}}}
{\mbox{exp}[\frac{q}{q+1}(E-\eta_2U^2)]}dU\nonumber\\
\hspace{-1cm}&=&F\left[
\frac{q}{q+1},\frac{q(\omega-2iq_1)}{2(q+1)\omega},
\frac{(3q+2)\omega-2iqq_1}{2(q+1)\omega},-\mbox{exp}\left[{\frac{i\omega U^2}
{(q+1)}}\right]\right]\nonumber\\
& &\times\frac{(q+1)^2\left[\sec\left[\frac{U^2\omega}{2(q+1)}\right]
\left(1+\mbox{exp}[\frac{i\omega U^2}{(q+1)}]\right)\right]^{\frac{q}{q+1}}}
{q(2(q+1)\eta_2+i\omega)}(\cosh[q_2]-\sinh[q_2]),
\end{eqnarray}
where $q_1=\eta_2(q+1)$, $q_2=\frac{q(E-\eta_2U^2)}{q+1}$ and $F$ is the hypergeometric function$^{\footnotesize{30}}$.
\section{Damped Mathews-Lakshmanan oscillator : Hamiltonian structure}
As a second example we consider
 the following damped version of the Mathews-Lakshmanan oscillator equation$^{\footnotesize{13}}$,
\begin{eqnarray}
\ddot{x}-\frac{\lambda_1x}{1+\lambda_1x^2}\dot{x}^2+ \frac{\alpha}
{1+\lambda_1x^2}\dot{x}
+\frac{\lambda x}{1+\lambda_1x^2}=0,\label{genmlo}
\end{eqnarray}
where we have included a nonlinear damping term to the Mathews-Lakshmanan oscillator.
The Mathews-Lakshmanan oscillator (equation (\ref{genmlo}) with $\alpha=0)$ possess simple
trigonometric solution$^{\footnotesize{13,14}}$, $x(t)=A\cos\Omega t$, $\Omega^2=\frac{\lambda}{1+\lambda_1A^2}$
with the Hamiltonian $H=p^2(1+\lambda_1x^2)+\frac{\lambda x^2}{1+\lambda_1x^2}$ .  We now find that the
damped case (\ref{genmlo}), with $\alpha\ne 0$ is also integrable and admits a conservative
Hamiltonian structure.

 Comparing the above equation with (\ref{lienard}), we find
\begin{eqnarray}
g(x)=\frac{1}{\sqrt{1+\lambda_1x^2}},\,f(x)=\frac{1}{(1+\lambda_1x^2)^{\frac{3}{2}}}.
\end{eqnarray}
The integral of motion is then
\begin{eqnarray}
I=\left\{
\begin{array}{ll}
\displaystyle\frac{(r-1)}{(r-2)}\frac{(r\dot{x}+\alpha x)}{r\sqrt{1+\lambda_1x^2}}
\left(\frac{r\dot{x}+\alpha x(r-1)}{r\sqrt
{1+\lambda_1x^2}}\right)^{1-r},&\alpha^2>4\lambda\\\\
\displaystyle\frac{2\dot{x}}{2\dot{x}-x}
-\log\left[\frac{2\dot{x}+\alpha x}{2\sqrt{1+\lambda_1x^2}}\right],&\alpha^2=4\lambda\\\\
\displaystyle\frac{\alpha}{2\omega}\tan^{-1}\left[\frac{\alpha
\dot{x}+2\lambda x}{2\omega\dot{x}}\right]
+\frac{1}{2}\log\left[\frac{\dot{x}^2
+\alpha x\dot{x}+\lambda x^2}{(1+\lambda_1x^2)}\right],&\alpha^2<4\lambda.
\end{array}
\right.
\end{eqnarray}
Substituting the above forms of $g(x)$ and $f(x)$ in the expression for the
Hamiltonian (\ref{lienard_ham}), we get
\begin{eqnarray}
\hspace{-1.8cm}H=\left \{
\begin{array}{ll}
\displaystyle\frac{(r-1)}{(r-2)}\left(p\sqrt{1+\lambda_1x^2}\right)^{\frac{r-2}{r-1}}
-\frac{\alpha(r-1)}{r}px,&\alpha^2>4\lambda\\\\
\displaystyle\frac{\alpha}{2}px+\log\left[p\sqrt{1+\lambda_1x^2}\right],& \alpha^2=4\lambda\\\\
\displaystyle\frac{1}{2}\log\left[\frac{x^2}{(1+\lambda_1x^2)}
\sec^2\left[\omega px\right]\right]
-\frac{\alpha p}{2}x,& \alpha^2<4\lambda,
\end{array}
\right.
\end{eqnarray}
where the canonically conjugate momentum
\begin{eqnarray}
p=\left \{
\begin{array}{ll}
\displaystyle\frac{1}{\sqrt{1+\lambda_1x^2}}
\left(\frac{r\dot{x}+\alpha x(r-1)}{r\sqrt{1+\lambda_1x^2}}\right)^{1-r},&\alpha^2\ge4\lambda\\\\
\displaystyle\frac{1}{\omega x}
\tan^{-1}\left[\frac{\alpha+2\dot{x}}{2\omega x}\right],&\alpha^2<4\lambda.
\end{array}
\right.
\end{eqnarray}
Using the canonical transformation $\displaystyle P=\frac{1}{2}\log(x^2),\,U=px$, one can
rewrite the Hamiltonian for the underdamped case ($\alpha^2<4\lambda$) as
\begin{eqnarray}
H=\frac{1}{2}\left(\log[\sec^2(\omega U)]-\log[e^{-2P}+\lambda_1]-\alpha U\right)\equiv E.
\end{eqnarray}
Following the same procedure illustrated in the previous example, one can arrive at the
following equation
\begin{eqnarray}
\dot{U}=1-\cos^2(\omega U)\exp[2E+\alpha U].
\end{eqnarray}
Obviously this can be  rewritten as the quadrature
\begin{eqnarray}
t-t_0=\int\frac{dU}{1-\cos^2(\omega U)\exp[2E+\alpha U]}.
\end{eqnarray}
The above integration can be easily performed for the choice $\alpha=0$ and one can
recover the known solution of the Mathews-Lakshmanan oscillator.  For $\alpha\ne0$, the
integration is nontrivial and requires further investigation.
Similarly, for the overdamped and the critically damped cases, one can use the canonical transformation
$U=\frac{1}{\sqrt{\lambda_1}}\sinh^{-1}(\sqrt{\lambda_1}x),\,\,P=p\sqrt{1+\lambda_1x^2}$ to reduce
their corresponding Hamiltonian to simpler forms.  However, again we are left with a quadrature for the choice $\alpha\ne0$.
Work is in progress to find other suitable
canonical transformations to find the general solution explicitly for the case $\alpha\ne0$.

\section{Lagrangian description}
We now present the equivalent Lagrangian description for the
systems studied so far.  One can follow two different procedures.
Since both the Hamiltonian, $H$, and the
corresponding canonical momentum, $p$, are available for each one of the systems studied
in the previous sections, one can write
down the associated Lagrangian as
\begin{eqnarray}
L=p\dot{x}-H.\label{lag_ham}
\end{eqnarray}
Alternatively, from the known form of the Lagrangian for the damped
harmonic oscillator given by us in Ref. 5, and applying the
nonlocal transformation (\ref{nlt}), one can deduce the
corresponding Lagrangian for the nonlinear damped oscillator
equations.  One can easily check that both the methods give rise
to the same Lagrangian.

Considering now the damped harmonic oscillator (\ref{damp}) and
its Hamiltonian form (\ref{damp_ham}) along with the conjugate
momentum (\ref{damp_mom}), the Lagrangian can be expressed as
\begin{eqnarray}
\hspace{-1.8cm}L=\left \{
\begin{array}{ll}
\displaystyle\left(y'+\frac{(r-1)}{r}\alpha y\right)^{(2-r)},&\alpha^2>4\lambda\\\\
\displaystyle\log\left[y'+\frac{1}{2}\alpha y\right],& \alpha^2=4\lambda\\\\
\displaystyle\frac{1}{2\omega}\left(\alpha\tan^{-1}\left[\frac{\alpha
y' +2\lambda y}{2\omega y'}\right] -\frac{2y'}{y}
\tan^{-1}\left[\frac{\alpha y+2y'}{2\omega y}\right]\right)\\
\displaystyle\qquad+\frac{1}{2}\log\left[y'^2+\alpha y'
y+\lambda y^2\right],& \alpha^2<4\lambda.
\end{array}
\right.\label{damp_lag}
\end{eqnarray}

Now applying the nonlocal transformation (\ref{nlt}) to the above
Lagrangian (\ref{damp_lag}), one can readily obtain the Lagrangian
associated with the general nonlinear damped oscillator equation
(\ref{lienard}) as
\begin{eqnarray}
\hspace{-1.8cm}L=\left \{
\begin{array}{ll}
\displaystyle\left(g(x)\dot{x}+\frac{(r-1)}{r}\alpha\int f(x)dx\right)^{(2-r)},&\alpha^2>4\lambda\\\\
\displaystyle\log\left[g(x)\dot{x}+\frac{1}{2}\alpha\int f(x) dx\right],& \alpha^2=4\lambda\\\\
\displaystyle\frac{\alpha}{2\omega}\tan^{-1}\left[\frac{\alpha g(x)\dot{x}
+2\lambda\int f(x)dx}{2\omega g(x)\dot{x}}\right]\\
\displaystyle\qquad-\frac{g(x)\dot{x}}{\omega\int f(x)dx}
\tan^{-1}\left[\frac{\alpha\int f(x)dx+2g(x)\dot{x}}{2\omega\int f(x)dx}\right]\\
\displaystyle\qquad+\frac{1}{2}\log\left[g(x)^2\dot{x}^2+\alpha g(x)\dot{x}\int f(x)dx+\lambda (\int f(x) dx)^2\right],& \alpha^2<4\lambda.
\end{array}
\right.\label{lienard_lag}
\end{eqnarray}
Naturally, the same form follows from the Hamiltonian
(\ref{lienard_ham}) and the conjugate momentum (\ref{lienard_mom})
using (\ref{lag_ham}). The Lagrangian for the specific examples
considered in Secs. IV and V can be deduced by specifying the
forms of $f(x)$ and $g(x)$ in equation (\ref{lienard_lag}).  It
may be noted that all the above Lagrangians are of non-standard
type, that is of the forms which cannot be written in the standard form as `kinetic energy' minus `potential energy'.  
In particular, one can readily check that the non-standard
Lagrangian
\begin{eqnarray}
L=\frac{1}{\dot{x}+\frac{2}{3}\int b(x) dx},
\end{eqnarray}
deduced by Musielak$^{\footnotesize{7}}$ for the nonlinear ODE
\begin{eqnarray}
\ddot{x}+b(x)\dot{x}+\frac{2}{9}b(x)\int b(x)dx=0,\label{musielak}
\end{eqnarray}
follows from the above general form (\ref{lienard_lag}) by
choosing $f(x)=b(x)$, $g(x)=1$,  $\alpha=1$ and $\lambda=\frac{2}{9}$ in equation (\ref{lienard}).  Note that the above equation (\ref{musielak}) itself
includes the MEE discussed in Refs. [8,9] as a special case with $b(x)=kx$.

\section{Conclusion}
In this paper, we have investigated a class of nonlinear dissipative systems which
admit a non-standard time independent Hamiltonian description through a novel nonlocal
transformation.  The procedure is simple
and straightforward.  By introducing a nonlocal transformation in the `source
equation', namely the DHO equation, we are able to generate a class of `target equations', namely
the nonlinear generalizations of DHO.  We have used the same nonlocal transformation
to deduce the time independent Hamiltonian for the nonlinear equation.  The
nonlocal transformation introduced in this paper is different from the one which
we have adopted in Ref. 22 and has certain salient features in identifying the
time independent Hamiltonian structure.
To illustrate the procedure we have
 considered two specific
systems, namely the generalized MEE and the damped
Mathews-Lakshmanan oscillator equation. We obtained the solution
of generalized MEE by integrating the canonical equations of
motion after introducing a suitable transformation.  The
associated Lagrangian description for all the above systems is
also briefly discussed.  A similar analysis can also be performed
in the case of two degrees of freedom systems. The details will be
presented separately.

\section{Acknowledgements}
The work forms a part of a research project of MS and an IRPHA
project of ML sponsored by the Department of Science \& Technology
(DST), Government of India.  ML is also supported by a DST Ramanna
Fellowship.

\begin{tabular}{p{.15cm}p{14cm}}

\footnotesize$^1$ &
P. G. Drazin \emph{Nonlinear Systems} (Cambridge University Press, Cambridge, 1992)\\
\footnotesize$^2$ &
G. Nicolis \emph{Introduction to Nonlinear Science} (Cambridge University Press, Cambridge, 1995)\\
\footnotesize$^3$ &
C. I. Um, K. H. Yeon, T. F. George \emph{ Phys. Rep.} {\bf 362} 63 (2002)\\
\footnotesize$^4$ &
H. Bateman \emph{Phys. Rev.} {\bf 38}, 815 (1931)\\
\footnotesize$^5$ &
V. K. Chandrasekar, M. Senthilvelan and M. Lakshmanan \emph{ J. Math. Phys.} {\bf 48},
032701 (2007)\\
\footnotesize$^6$ &
V. K. Chandrasekar, M. Senthilvelan and M. Lakshmanan \emph{J. Phys. A: Math. Theor.}
{\bf 40}, 47171 (2007)\\
\footnotesize$^7$ &
Z. E. Musielak  \emph{J. Phys. A : Math. Theor.} {\bf 41}, 055205 (2008)\\
\footnotesize$^8$ &
J. F. Carinena and M. F. Ranada \emph{J. Math. Phys.} {\bf 46}, 062703 (2005)\\
\footnotesize$^9$ &
V. K. Chandrasekar, M. Senthilvelan and M. Lakshmanan \emph{Phys. Rev. E} {\bf 72},
066203 (2005)\\
\footnotesize$^{10}$  &
M. R. Feix, C. Geronimi, L. Cairo, P. G. L. Leach, R. L. Lemmer
and S. E. Bouquet \emph{J. Phys. A Math. Gen.} {\bf 30}, 7437 (1997)\\
\footnotesize$^{11}$  &
S. E. Bouquet, M. R. Feix and P. G. L. Leach \emph{J. Math. Phys.} {\bf
32}, 1480 (1991)\\
\footnotesize$^{12}$  &
V. K. Chandrasekar, M. Senthilvelan and M. Lakshmanan \emph{ J. Phys. A: Math. Gen.}
{\bf 39}, L69 (2006)\\
\footnotesize$^{13}$  &
P. M. Mathews and M. Lakshmanan  \emph{Quart. Appl. Maths.} {\bf 32}, 315 (1974)\\
\footnotesize$^{14}$  &
M. Lakshmanan and S. Rajasekar \emph{Nonlinear Dynamics : Integrability, Chaos and Patterns}
(Springer-Verlag, New York, 2003)\\

\footnotesize$^{15}$  &
E. L. Ince \emph{Ordinary Differential Equations} (Dover, New York, 1956)\\

\footnotesize$^{16}$  &
H. T. Davis \emph{Introduction to Nonlinear Differential and
Integral Equations} (Dover, New York, 1962)\\
\footnotesize$^{17}$  &
I. C. Moreira  \emph{Hadronic. J} {\bf 7}, 475 (1984)\\
\footnotesize$^{18}$  &
 P. G. L. Leach \emph{J. Math. Phys.} {\bf 26}, 2510 (1985)\\
\footnotesize$^{19}$  &
V. J. Erwin, W. F. Ames and E. Adams \emph{Wave Phenomena: Modern Theory
and Applications} ed C. Rogers and J. B. Moodie (Amsterdam, 1984)\\
\footnotesize$^{20}$  &
S. Chandrasekhar  \emph{An introduction to the Study of Stellar Structure}
(Dover, New York, 1957)\\
\end{tabular}

\begin{tabular}{p{.15cm}p{14cm}}

\footnotesize$^{21}$  &
V. K. Chandrasekar S. N. Pandey, M. Senthilvelan and M. Lakshmanan  \emph{J. Math.
Phys.} {\bf 47}, 023508 (2006)\\
\footnotesize$^{22}$  &
V. K. Chandrasekar, M. Senthilvelan, A. Kundu and M. Lakshmanan \emph{J. Phys. A. Math. Gen.} {\bf 39},
9743 (2006)\\

\footnotesize$^{23}$  &
P. Kustaanheimo and E. Stiefel  J Reine  \emph{Angew. Math.} {\bf 218}, 204 (1965)\\

\footnotesize$^{24}$  &
G. M. Murphy \emph{Ordinary Differential Equations and Their Solutions}
(Affilitated East-West Press, New Delhi, India, 1960)\\

\footnotesize$^{25}$ &
R. Iacono \emph{J. Phys. A : Math. Theor.} {\bf 41}, 068001 (2008)\\
\footnotesize$^{26}$ &
V. K. Chandrasekar, M. Senthilvelan and M. Lakshmanan \emph{J. Phys. A: Math. Theor.}
{\bf 41}, 068002 (2008)\\
\footnotesize$^{27}$  &
V. K. Chandrasekar, M. Senthilvelan and M. Lakshmanan \emph{Proc. R. Soc. London,
Soc. A }{\bf 461}, 2451 (2005)\\
\footnotesize$^{28}$  &
M. A. Reyes and H. C. Rosu \emph{ J. Phys. A : Math. Theor.} {\bf 41}, 285206 (2008)\\
\footnotesize$^{29}$  &
A. R. Janzen  \emph{Private Communication } (2007)\\
\footnotesize$^{30}$  &
I. S. Gradshteyn and I. M. Ryzhik
\emph{Table of Integrals, Series and Products}
(Academic Press, London, 1980)

\end{tabular}


\end{document}